\documentclass[aps,prl,reprint,superscriptaddress]{revtex4-2}

\usepackage{graphicx}
\usepackage{textcomp}
\usepackage{textgreek}
\usepackage{amsmath}
\usepackage{gensymb}
\usepackage{caption}
\begin{document}


\title{Optical spin pumping in silicon}


\author{Stefano Achilli}
\affiliation{Dipartimento di Scienza dei Materiali, Universit\`a degli Studi di Milano-Bicocca and BiQuTe, Via R. Cozzi 55, 20125 Milano, Italy}

\author{Damiano Marian}
\affiliation{Dipartimento di Fisica, Universit\`a di Pisa, Largo Pontecorvo 3, I-56127 Pisa, Italy}

\author{Mario Lodari}
\affiliation{QuTech and Kavli Institute of Nanoscience, Delft University of Technology, 2600 GA Delft, Netherlands.}

\author{Emiliano Bonera}
\affiliation{Dipartimento di Scienza dei Materiali, Universit\`a degli Studi di Milano-Bicocca and BiQuTe, Via R. Cozzi 55, 20125 Milano, Italy}

\author{Giordano Scappucci}
\affiliation{QuTech and Kavli Institute of  Nanoscience, Delft University of Technology, 2600 GA Delft, Netherlands.}

\author{Jacopo Pedrini}
\affiliation{Dipartimento di Scienza dei Materiali, Universit\`a degli Studi di Milano-Bicocca and BiQuTe, Via R. Cozzi 55, 20125 Milano, Italy}

\author{Michele Virgilio}
\affiliation{Dipartimento di Fisica, Universit\`a di Pisa, Largo Pontecorvo 3, I-56127 Pisa, Italy}

\author{Fabio Pezzoli}
\email[]{fabio.pezzoli@unimib.it}
\affiliation{Dipartimento di Scienza dei Materiali, Universit\`a degli Studi di Milano-Bicocca and BiQuTe, Via R. Cozzi 55, 20125 Milano, Italy}



\begin{abstract}
The generation of an out-of-equilibrium population of spin-polarized carriers is a keystone process for quantum technologies and spintronics alike. It can be achieved through the so-called optical spin orientation by exciting the material with circularly polarized light. Although this is an established technique for studying direct band-gap semiconductors, it has been proven limited in materials like Si that possess weak oscillator strengths for the optical transitions. In this study, we address the problem by presenting an all-optical analog of the spin pumping method. This involves the optical creation of a non-equilibrium spin population within an absorber, which subsequently transfers spin-polarized carriers to a nearby indirect gap semiconductor, resulting in polarized emission from the latter. By applying this concept to a Ge-on-Si heterostructure we observe luminescence from Si with an unrivaled polarization degree as high as 9\%. The progressive etching of the absorbing layer, assisted by magneto-optic experiments, allows us to ascertain that the polarized emission is determined by effective spin injection aided by the carrier lifetime shortening due to extended defects. These findings can facilitate the use of highly promising spin-dependent phenomena of Si, whose optical exploitation has thus far been hampered by fundamental limitations associated with its peculiar electronic structure.
\end{abstract}


\maketitle

\section{Introduction}\label{sec1}
The ability to inject and detect spins in solid-state systems is of prime scientific and technological interest. Light-matter interaction in the presence of spin-orbit coupling has been shown to be one of the most effective  strategies to achieve this purpose.\cite{lampel68} In semiconductors, the absorption of circularly polarized light transfers angular momentum to charge carriers, generating an out-of-equilibrium spin ensemble: a process commonly refereed to as optical spin orientation.\cite{Dyakonov_OO, zutic04} The spontaneous evolution of the system towards thermal equilibrium can then occur through radiative recombination events. Due to the time reversal symmetry of the selection rules, spin-dependent optical transitions yield emission of circularly polarized light, making photoluminescence (PL) a convenient experimental tool for fast and contactless detection of the spin properties.\cite{Dyakonov_OO, zutic04}

This all-optical injection-detection scheme was originally applied to direct-gap semiconductors, such as GaSb \cite{Parsons69} or GaAs, \cite{Dyakonov_OO} and lately contributed to gather a deeper understanding of the spin physics of emerging systems such as two-dimensional materials \cite{cao12, Sallen2012} and perovskites. \cite{Odenthal2017, lu22} Despite all these advances, optical spin investigation continues to be highly ineffective for a large class of materials, namely indirect gap semiconductors. The latter are jeopardized by the weak phonon-mediated transitions that make spin-dependent absorption and recombination inherently inefficient processes. The specific magnitude of the oscillator strength for these dipole-allowed transitions indeed results in (i) a weak spin-injection rate, washing out almost completely the average electron spin along the light propagation direction and (ii) a lengthening of the lifetime of the electrons in the fundamental indirect valleys to such an extent that spin relaxation occurs on a time scale faster than the recombination process, yielding, in other words, unpolarized PL. 

\begin{figure}
\centering
  \includegraphics[width=\linewidth]{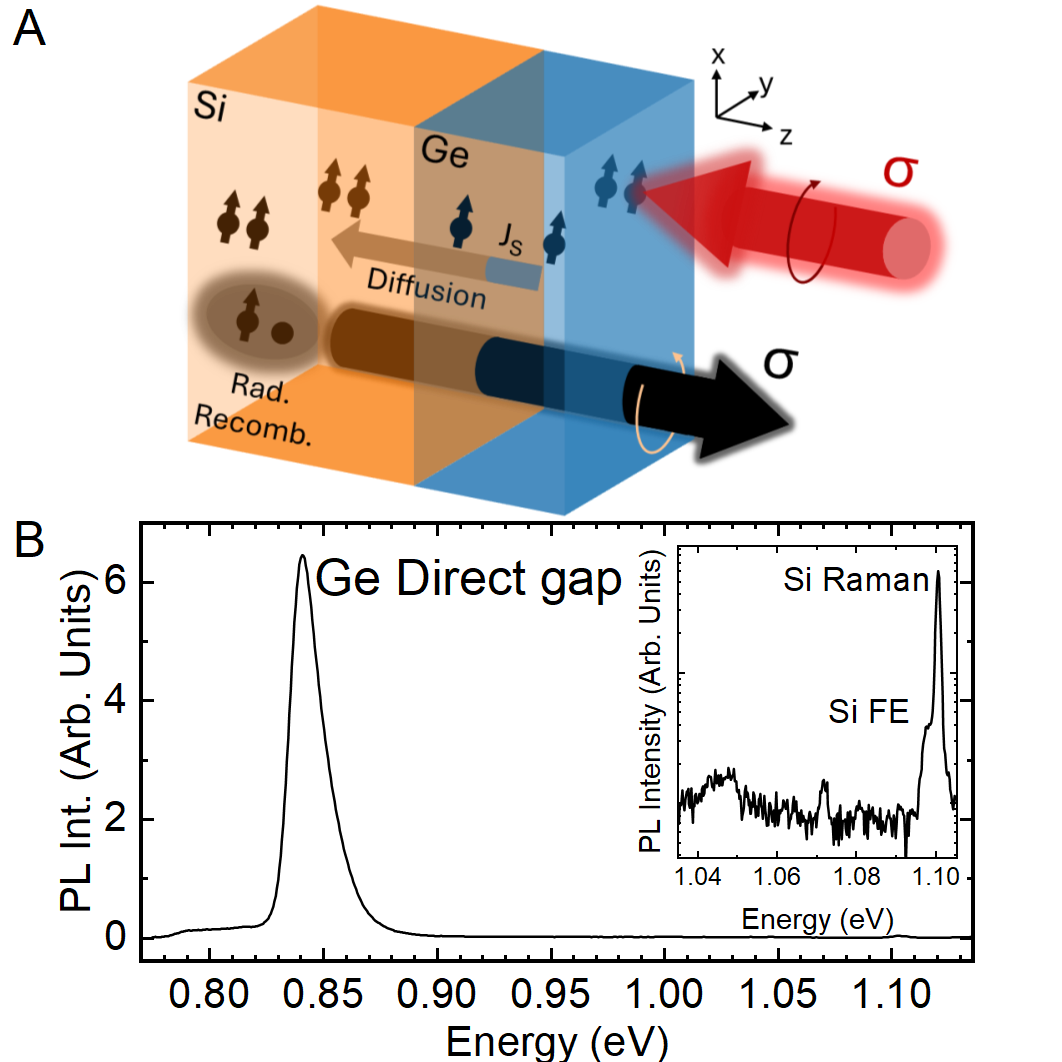}
  \caption{\textbf{Schematics of the optical spin pumping effect.} \textbf{(A)} Ge acts as spin injector under optical excitation with circularly polarized light ($\sigma$, shown in red). Si functions as a spin sink and emits polarized luminescence (shown in black). The electron spin current $\mathrm{J_S}$ flows from Ge to Si because of diffusion gradient and thanks to the favorable band alignment. \textbf{(B)} Photoluminescence spectrum of the Ge/Si heterostructure measured at T=4 K under excitation with a 1.17 eV laser showing that the dominant peak is due to direct gap emission from the Ge epilayer. The inset shows a magnification at higher resolution in the spectral region of the Si band gap, highlighting from low to high energy: (i) low intensity processes like the overtone Raman scattering or the recombination involving multiple emission of zone center phonons and intervalley scattering (below $\approx$1.09 eV), (ii) the free exciton (FE) recombination involving transverse-optical (TO) phonons (at 1.097 eV) and (iii) the one-phonon Raman mode ($\approx$1.1 eV), respectively.}
  \label{fig:1}
\end{figure}

The most prominent example in this context is Si, where there has been long-awaited experimental evidence for circularly polarized emission.\cite{zutic06, nastos07} In this material, not only resonant excitation at the indirect gap ($\sim 1.1$ eV) is hindered by the subtleties of phonon-assisted transitions, but a complete spin relaxation is expected to make unpractical also the excitation through the remote direct gap ($\sim 3.4$ eV).\cite{cardona66} Various methods using highly polarizing magnetic fields, \cite{jonker07, sircar14} electro-optical hybrids, \cite{bottegoni17, Zucchetti23} or purely electrical/thermal means \cite{Grenet09, dash09, Spiesser17, le_breton11} have been pursued as attempts to cope with the unfeasible, although more direct, measurement of the weak luminescence polarization. A recent work sets indeed the fundamental detection limit for the degree of polarization under convenient continuous-wave excitation to unpractical vanishing values of $\sim 10^{-4}$, \cite{marie25} demonstrating that in Si the all-optical approach still remains a severe issue. Such a finding is in stark contrast with the fact that Si enabled the first experimental demonstration of optical spin injection in the solid state, \cite{lampel68} and that the theoretical framework for such a process is now well understood. \cite{nastos07,li10,li13} 

In this work, we propose a distinct approach that circumvents the peculiar limitations of using optics to study indirect gap materials. We exemplarily apply this method to a prototypical semiconductor such as Si. Specifically, we mimic the technique of spin pumping \cite{ando12, Shikoh13} by introducing the optical analog depicted in the schematics of Fig. \ref{fig:1}. Spin pumping relies on radiofrequency-induced magnetization dynamics to drive a spin current ($\mathrm{J_S}$) across a semiconductor/ferromagnet interface. \cite{Tserkovnyak05} Here, instead, we apply optical orientation to excite a macroscopic magnetic moment and initiate spin dynamics in a light-absorbing medium. The latter replaces the ferromagnetic injector with a semiconducting film, which is adjacent, in our case, to Si and whose direct gap transitions must be suitably accessible. Inspired by a materials-by-design approach, our choice for the spin generator naturally falls on Ge. In addition to solving practical manufacturing constraints related to the epitaxy on Si and enabling integration in electronic devices, Ge offers highly polarized direct gap excitation \cite{Loren09, Rioux10, li13, Pezzoli13} and a convenient band alignment at the interface with Si. \cite{kuo05, Virgilio06} The band lineup spontaneously results in diffusion and transfer of spin-polarized electrons towards Si, even in the complete absence of an external electrical bias and without any impedance mismatch issue. This seemingly guarantees straightforward accumulation of a thermalized spin population directly in the Si sink, favoring radiative recombination events. Such profitable mechanisms, together with a shortening of the electron lifetime at the heterojunction, boast almost five orders of magnitude improvement of the circular polarization degree with a best-in-class value as high as 9\%.\cite{marie25} The observation of the optical Hanle effect and a numerical model of the carrier photogeneration and diffusion process further corroborate such distinct result. 
The proposed approach can drastically widen the access to intriguing spin-related properties of Si, whose optical exploitation has been hitherto narrowed due to the fundamental indirect nature of its electronic structure.\cite{dery11, jansen12} More generally, these findings have the potential to open novel research directions in achieving spin injection and detection in spintronics and quantum technologies, harnessing materials that are technologically relevant and at the core of modern manufacture of electronic and photonic circuits.

\section{Excess carrier dynamics in Ge/Si heterostructure}\label{sec2}

Figure \ref{fig:1}B shows a PL spectrum of a heterostructure composed of a Ge film epitaxially grown on a (001)Si wafer (see methods for details). The measurement is carried out at a lattice temperature of 4 K using a Nd:YVO$_4$ laser at 1.165 eV to leverage the absorption through the direct gap of the epilayer. The spectrum is indeed dominated by a peak at $\sim$0.84 eV, pertaining to the interband transitions in Ge occurring at the zone center, i.e., the Ge direct gap.\cite{Pezzoli13, de_cesari18} Interestingly, in the energy range above the Ge band edge, at about $\sim$1.1 eV, an additional weaker signal can be resolved. A high-resolution measurement demonstrates the presence of various components. The most prominent one consists of a relatively brighter peak at 1.101 eV. Its energy separation from the laser line and its temperature-dependent behavior (see Supplementary Materials) are consistent with the well-known one-phonon Raman spectrum of Si due to the zone-center longitudinal optical (LO) phonons. The magnified spectrum in Figure \ref{fig:1}B shows a set of minor bands observed at slightly lower energy, which are the superposition of second-order phonon Raman scattering \cite{WANG73, Temple73} and low-intensity radiative recombinations.\cite{Dean67}  

Notably, the intense one-phonon Raman mode is in close proximity and superimposed to a weaker structure that appears as a red-shifted shoulder at 1.097 eV. Such energy value is a strong indication that the likely source for this feeble component is the recombination of free excitons (FE) that develop, also in this case, from the Si substrate.\cite{Colley87,davies89} This spectral line originates from phonon-mediated radiative recombination involving transverse-optical (TO) phonons, which is known to dominate the low-temperature PL spectrum of lightly-doped unstrained Si.\cite{Colley87,davies89} The corresponding replica assisted by LO phonons is relatively weaker and blue-shifted, hence potentially concealed by the one-phonon Raman peak. The formation of electron-hole droplets associated with the characteristic broad lower-energy emission has not been observed under our excitation conditions, pointing towards a low-injection regime.\cite{davies89} The electronic nature of the FE peak is further confirmed by the observation of a Varshni-like temperature-induced behavior (see Supplementary Materials). These experimental evidences openly suggest that the PL emission from the buried Si substrate accompanies the direct photoexcitation of carriers in the topmost Ge film. 

\begin{figure*}
\centering
\includegraphics[width=\linewidth]{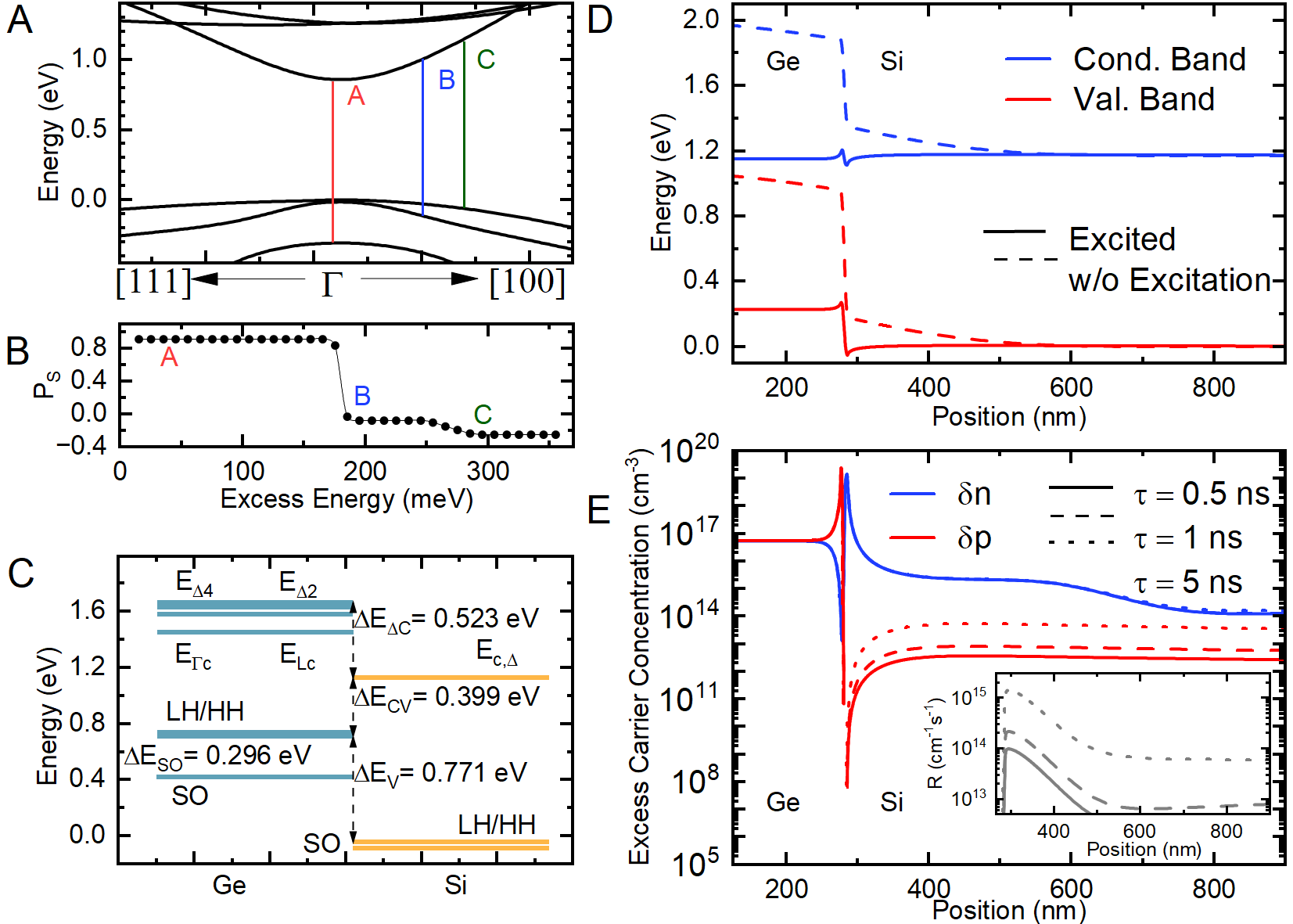}
\caption{\textbf{Excess carrier modeling.} \textbf{(A)} Low temperature electronic spectrum of Ge under biaxial tensile strain of 0.2\%. \textbf{(B)} Electronic spin polarization $P_s$ as a function of the excess kinetic energy with respect to the bottom of the conduction band edge $\Gamma_c$. \textbf{(C)} Valence and conduction band edge alignments between cubic Si (yellow) and 0.2\% strained Ge (light blue). \textbf{(D)} Valence (red) and conduction (blue) self-consistent band edge profiles for the n-type Si / p-type Ge heterojunction without (dashed) and with (solid) illumination (3 kW/cm$^2$) for $\tau_{Si}=1$ ns. In panel \textbf{(C)} and \textbf{(D)} the reference energy is kept constant at the Si HH level. \textbf{(E)} Excess carrier concentration profiles $\delta p$ (red) and $\delta n$ (blue) in a neighbor of the heterointerface, calculated for different values of $\tau_{Si}$. The inset reports the corresponding recombination rate $R$.}
\label{figTeo}
\end{figure*}

In order to fruitfully utilize such finding to achieve spin-polarized optical recombination from Si, we begin by modeling the band structure around the $\Gamma$ point of (001)Ge, considering a 0.2\% biaxial strain caused by epitaxial growth at high temperatures.\cite{pezzoli08, Pezzoli08b, Sammak19} The electronic spectrum shown in Figure \ref{figTeo}A has been calculated by means of first neighbor tight-binding Hamiltonian, with sp$^3$d$^5$s$^*$-orbitals in both spin configurations, as detailed in Ref. \cite{Vitiello15}. A tensile strain field is known to be induced upon cooling the Ge/Si heterostructure to cryogenic temperatures, due to the mismatch between the lattice thermal expansion coefficients of Si and Ge. This results in a splitting of the degeneracy of the light-hole (LH) and heavy-hole (HH) levels at the $\Gamma$ point and in a reduction of the amplitude of the direct gap. It should be noted that strain has been included in the model to provide a more realistic description of the heterosystem, even though its relatively small value has negligible impact on the overall spin and charge dynamics. 

At the chosen pump energy, the vertical transition involving the split-off (SO) bands is active, as schematically depicted by the red arrow in Fig. \ref{figTeo}{A}. Those photo-excited electrons feature very small excess energy with respect to the conduction band edge,  $\Gamma_{c}$. By virtue of the optical selection rules, a normally incident $\sigma^{+}$ circularly polarized excitation is expected to result for these SO-coupled electrons in a net spin polarization pointing along the quantization axis (Fig. \ref{figTeo}{B}). The latter is defined by the light propagation direction and corresponds with the growth direction, $z$. Due to energy conservation, electrons originating from the SO band, hence sitting close to the bottom of the $\Gamma_c$ valley, are expected to provide a negligible contribution to the Si FE shown in Fig. \ref{fig:1}B. Rather, they primarily concur to the Ge recombination either through the direct gap transitions or $L$-valley indirect emission. The latter occurs upon intervalley scattering.

It is worth noting that the optical pump also excites conduction electrons from HH and LH bands (see arrows C and B in Fig. \ref{figTeo}{A}. As shown in Fig. \ref{figTeo}{B}, this electron population has the largest excess energy within the $\Gamma_c$ ensemble and, at the same time, sustains a spin polarization ($P_{s}$) as high as $\approx 0.3$. This value originates from the modest energy separation between the two valence bands and from the mixing of the total angular momentum $J$, induced by the finite value of the wave-vector $k$. We therefore argue that such spin-polarized high-energy electrons can play a crucial role in the spin-pumping process.

To demonstrate that their intertwined spin and energy relaxation dynamics can eventually yield circularly polarized light also from the Si region, we show in Fig. \ref{figTeo}{C} the valence (HH,LH,SO) and conduction (L,$\Gamma_c$,$\Delta$) flat-band alignments between cubic Si and tensile-strained Ge, calculated following Refs. \cite{Walle1986,Walle1989}.
After photoexcitation in a neighbor of $\Gamma_c$ of Ge, high-energy electrons can subsequently diffuse towards the heterojunction, while experiencing phonon-induced intervalley scattering towards the $\Delta$- rather than the $L$-valleys, owning to a larger transition rate. \cite{Pezzoli13} Thanks to this process, these electrons can be eventually funneled into Si due to a favorable band lineup (type II) with the $\mathrm{\Delta^{Si}}$ band edge. In other words, an electron ensemble can be injected, virtually thermalized, at the absolute minimum of the Si conduction band. This can be particularly beneficial for enhancing the possibility of observing polarized PL, as it minimizes spin losses associated to energy relaxation and dwelling in the Si conduction band. 

It should be also noticed that the robust potential barriers present at both the SO and HH/LH band edge profiles retain holes on the Ge side of the interface. 
It follows that the PL observed in Figure \ref{fig:1}B at the Si FE energy of 1.097 eV, requires holes to be generated directly in the Si region. Direct photo-excitation in Si is indeed possible, although the resulting excess carrier concentration must be relatively small, since the absorption coefficient at the pump energy is modest, i.e., of the order of 10 cm$^{-1}$, and the type II band lineup tends to drain holes out of Si.

To quantitatively assess the excess carrier concentration profile, we have performed self-consistent drift-diffusion simulations considering the optically excited Si/Ge heterointerface in a two-band approximation, assuming that the large majority of electrons and holes populate the $\Delta$ valley and the HH band, respectively. 
To this purpose, we included both direct and indirect optical absorption of the two materials, the related pump depletion effects, and considering n- and p-type doping for Si and Ge, respectively (see methods). Pump reflection at the Si/Ge interface has been also implemented in the calculation. As for the recombination dynamics, we expect the excess carrier lifetime being dominated by a Shockley-Read-Hall mechanism, due to the well-known presence of lattice defects at the epitaxial interface.\cite{Geiger14, Pezzoli14, Pezzoli16} Thus, we have modeled this recombination channel following Ref. \cite{RSH} setting the same recombination time $\tau$ for electrons and holes. Due to the large uncertainty affecting the value of this materials parameter, which is strongly excitation and sample dependent, the simulations have been performed varying $\tau_{Si}$ in the Si region in the 0.5-5 ns range, while keeping a fixed value $\tau_{Ge}$ of 1 ns for the defective Ge layer.\cite{Geiger14} 
Since the presence of a type-II band discontinuity is accompanied by electrostatic effects due electron-hole charge displacement, the Hartree interaction potential has been taken into account in a self-consistent iterative scheme. A modified Scharfetter-Gummel discretization approach has been adopted to implement the Fermi-Dirac statistics. \cite{ScharfetterGummel,Lundstrom1983} 
Materials parameters are taken from the literature and are listed in the Supplementary Materials (Table S1).

In Figure \ref{figTeo}{D} we show the band-edges at equilibrium (i.e., without illumination) for a typical sample with a Ge thickness of 280 nm. Using these parameters and the doping levels characterizing the two semiconductors forming the junction (see methods), the type II p-n heterojunction almost approaches a broken-gap condition, resulting in an energetic barrier for holes that is practically equal to the Si gap. Consequently, spin-polarized electrons experience a potential step roughly equal to the Ge gap. 

When the sample is optically excited, with a pump fluency of, e.g., 3 kW/cm$^2$, a large density of carrier is created in the Ge region, resulting in a negative charge transfer to the Si material, triggered by the aforementioned potential step.
Remarkably, the hole backflow, i.e., from Si to Ge, is much weaker because of the significantly smaller value of the absorption coefficient in Si with respect to Ge. 
As a result, it can be observed a complete flattening of the conduction band edge profile and a lowering of the valence band discontinuity, whose value is now close to the gap difference between Si and Ge.

The corresponding excess electron (hole) density $\delta n= n - n_{eq}$ ($\delta p = p - p_{eq}$) is reported in Figure \ref{figTeo}E, for $\tau_{Si}$ equal to 0.5, 1, 5 ns. Considering that bulk Si features a much larger excess carrier lifetime, this choice is motivated by the presence of crystal defects near the Ge interface acting as an efficient non-radiative recombination channel (see the following Section discussing magneto-optical experiments).
As expected, away from the Si/Ge junction the excess electron and hole concentration profiles flatten and tend to a common value (not shown in the Figure for Si). 
Well-inside the Si region this quantity is controlled by the indirect absorption of the pump and varies in the $10^{12}-10^{13}$ cm$^{-3}$ range, depending on the chosen $\tau_{Si}$.
Similarly, close to the top surface, the valence and conduction excess carriers have practically the same concentration, which for $\tau_{Ge} = 1$ ns is about $5 \cdot 10^{16}$ cm$^{-3}$, due to the much larger optical absorption of Ge.
Close to the interface, we observe an accumulation (depletion) of conduction carriers in the Si (Ge) region, with peak values exceeding $10^{18}$ cm$^{-3}$, triggered by the electron transfer from the Ge material and the attractive electrostatic interaction with holes confined in the epilayer.
The same effect controls the complementary sharp variation of hole concentration across the interface.

Using the carrier concentration profiles, we have estimated the spatially-resolved radiative recombination rate $R=B(np-n_{eq}p_{eq})$, reported in the inset of Fig. \ref{figTeo}E by varying $\tau$. Here B is the radiative recombination coefficient (see Supplementary Materials).
We find that $R$ peaks at the interface, with values ranging from 10$^{14}$ to 10$^{15}$ s$^{-1}$cm$^{-3}$, decaying inside the Si material with a length scale of $\sim100$ nm. For comparison, we observe that similar $R$ peak values in intrinsic bulk Si at low T require an excess carrier concentration $\delta p=\delta n\sim10^{14}-10^{15}$ cm$^{-3}$. Finally, we have performed numerical experiments to consider also a spatial inhomogeneous crystal quality in Si, related to the quenching of the defects density far away from the Ge/Si interface. To this end, we have set larger excess carrier lifetimes for depths $\geq 400$ nm. In this case, the main effect is the enhancement of the hole density close to the Si top surface, which is caused by the larger diffusion length of unpolarized holes generated deep inside the Si bulk. This results in a overall increase of the spatially-resolved recombination rate, which, nevertheless, still peaks within 100 nm from the heterojunction. 

In conclusion, the calculations indicate that the PL observed at the Si FE energy can be safely attributed to the optical recombination of electrons, originally excited in Ge by optical transitions from the HH/LH bands. As a matter of fact, these carriers efficiently diffuse in Si where they recombine radiatively with holes created therein by pump absorption through the indirect gap. Despite the small hole concentration in Si, the large electronic density due to carrier accumulation at the Si/Ge type II interface, results in observable recombination rates $R$ in the Si region in the proximity of the heterointerface.

This interesting mechanism leads to other notable conclusions. Electrons originating from HH/LH states can be remarkably polarized through optical orientation (see Fig. \ref{figTeo}). The long-lived electron spins in Ge guarantee sizable spin diffusion along with reduced intervalley losses. \cite{Giorgioni14, giorgioni16} It is therefore reasonable to expect that optical spin pumping can take place in the heterostructure with Ge acting as an electron spin injector and Si serving as a spin sink. Even tough photogenerated holes will be almost unpolarized given their relatively short spin relaxation time, quite naturally radiative recombination events will involve the accumulated electron spins, potentially leading to the observation of a robust and sought-after circularly polarized PL from Si. 

\begin{figure*}
\centering
  \includegraphics[width=0.75\linewidth]{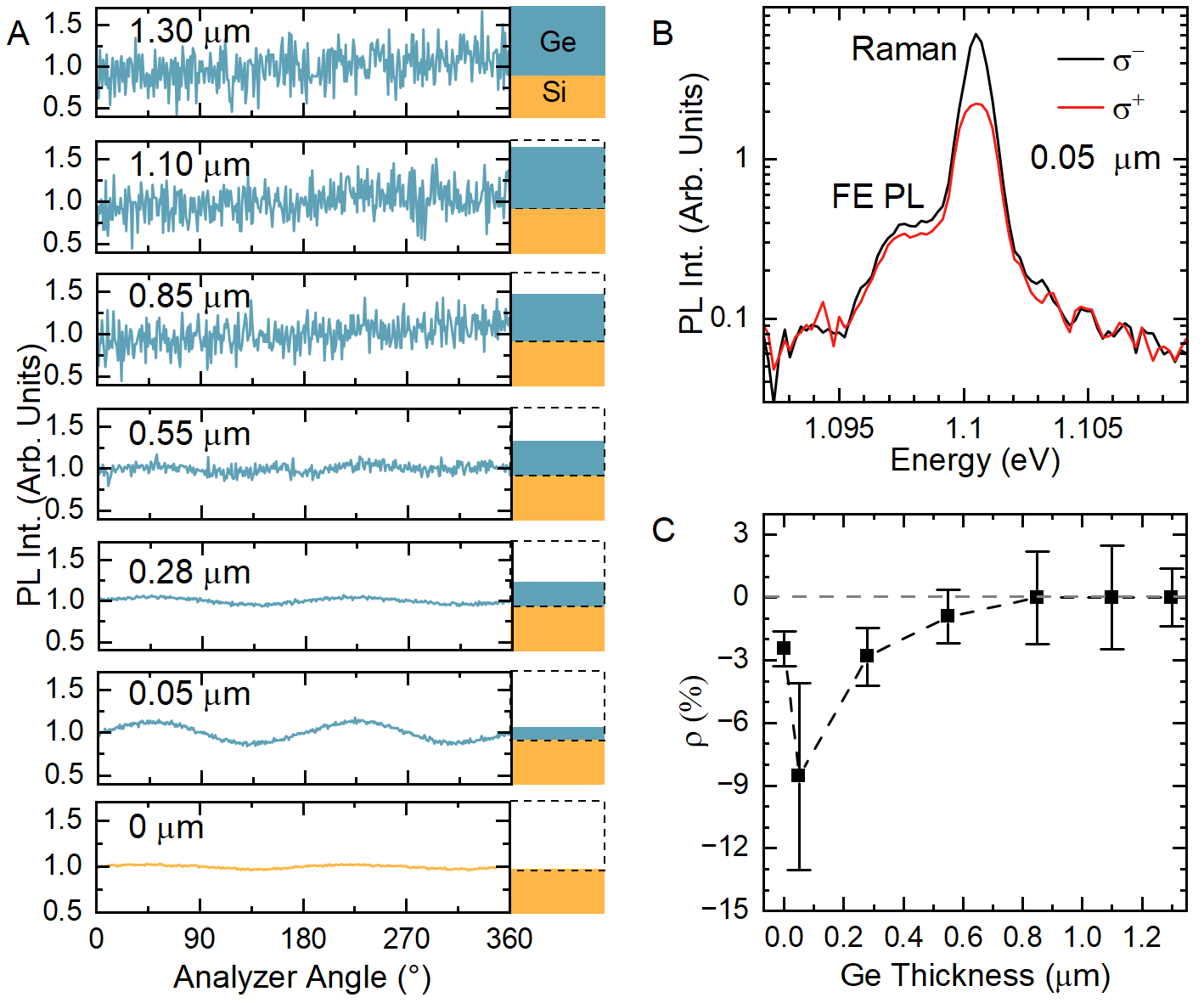}
    \caption{\textbf{Observation of polarized emission from Si.} \textbf{(A)} Polarization analysis of the Si FE PL at 1.097 eV performed at T=4 K as a function of the relative orientation (analyzer angle) between an optical retarder and a polarizer. The investigation is performed for various Ge thicknesses by chemically etching the Ge film from 1.3 $\mu$m down to 0 $\mu$m. The latter corresponds to a complete removal of the Ge layer. \textbf{(B)} Polarization-resolved PL components of the Si FE measured at T=4 K for a Ge thickness of 0.05 $\mu$m. The black curve corresponds to the left-handed, $\sigma^-$, component of the emitted light, whereas the red line shows the right-handed, $\sigma^+$, component. \textbf{(C)} Polarization degree $\rho$ as a function of the Ge film thickness. The error bars are obtained from the standard deviation of several measurements.}
  \label{fig3}
\end{figure*}

\section{Experimental demonstration of the optical spin pumping effect}

To leverage the convenient band-alignment and demonstrate the concept of optical spin pumping shown in Figure \ref{fig:1}A, we performed optical spin orientation in Ge, while conducting an analysis of the polarization pertaining to the Si FE emission. Specifically, the degree and state of polarization of the PL can be comprehensively characterized by measuring the modulation of the amplitude of PL peaks as a function of the relative orientation (analyzer angle) between an optical retarder and a polarizer.\cite{Goldstein, pezzoli12, Pezzoli13} 

This investigation was repeated whilst the Ge film was progressively thinned by selective wet etching until its full removal.\cite{Huygens2007} The outcomes, summarized in Fig. \ref{fig3}A, unveil an intriguing scenario. The polarization analysis of the Si FE performed on the pristine Ge/Si heterostructure (i.e., Ge thickness $1.3$ $\mathrm{\mu m}$) shows indeed random fluctuations of the PL intensity, thus suggesting that only a minor fraction of the spin-oriented carriers is likely to be injected into Si. As a consequence, the polarized emission, if present, remains well below the noise level of the PL, akin to the conventional direct excitation of bulk Si. Such condition persists while further reducing the Ge thickness. Remarkably, however, as the thickness of the epitaxial layer approaches 0.55 $\mu$m, a net modulation of the PL starts to appear, revealing a sinusoidal behavior whose amplitude is enlarged by the subsequent etching steps. When the thickness of the Ge layer is reduced to 0.05 $\mu$m, the polarization curve openly manifests a distinct $90 \degree$ periodicity oscillation with the largest peak-to-valley ratio. The latter is the hallmark of a significant circular polarization degree of the emission and is ultimately related to the optically-oriented electron spins. This finding can also be better appreciated by having a close look at the corresponding PL spectra resolved in Figure \ref{fig3}B for the $\sigma^-$ and $\sigma^+$ polarizations. The former component is clearly larger than the latter both for the Raman mode and, most importantly, the Si FE. 

The transition from a featureless behavior to a full periodic modulation of the PL manifests itself also in the degree of circular polarization $\rho$. Figure \ref{fig3}C summarizes the statistical values retrieved from the polarization analysis conducted on several measurements on each sample. Initially, when the Ge layer is thick, the PL is not polarized, so $\rho$ is null. However, once the thickness of the Ge absorbing layer decreases, $\rho$ becomes increasingly negative. All these evidences make a compelling case for the emergence of circularly polarized PL from Si. Moreover, in striking agreement with the mechanism of optical spin pumping anticipated previously, the sign of the circular polarization of the Si FE remarkably demonstrates an opposite helicity with respect to the one pertaining to the Ge direct gap (see Supplementary Materials Fig. 2). Indeed, the latter emission is governed by electrons optically coupled to the SO states. The spin orientation of such population is antiparallel to the one of electrons photoexcited from the top of the valence band. Such carriers suffer intervalley scattering, which steers them out of the zone center and, following diffusion, ultimately gives rise to the Si FE and to its polarization. 

By further etching the sample, carriers are no longer generated in Ge, but rather in Si, and the Si polarization fades away as demonstrated in Fig \ref{fig3}C. Notice also the suppression of the amplitude oscillations in Fig. \ref{fig3}A. Remarkably, $\rho$ reaches a minimum when the Ge film is 0.05 $\mu$m thick with an absolute value $\mid\rho\mid$=9\% that is almost five orders of magnitude larger than what can be expected for resonant excitation at the steady state in an indirect material such as Si.\cite{marie25} We can therefore argue that the mechanisms of optical spin pumping and spin accumulation introduced before are indeed at play and outperform conventional excitation schemes by providing a key enabling strategy for injecting a giant spin polarization in Si and, more generally, in indirect gap materials.

\begin{figure*}
\centering
  \includegraphics[width=\linewidth]{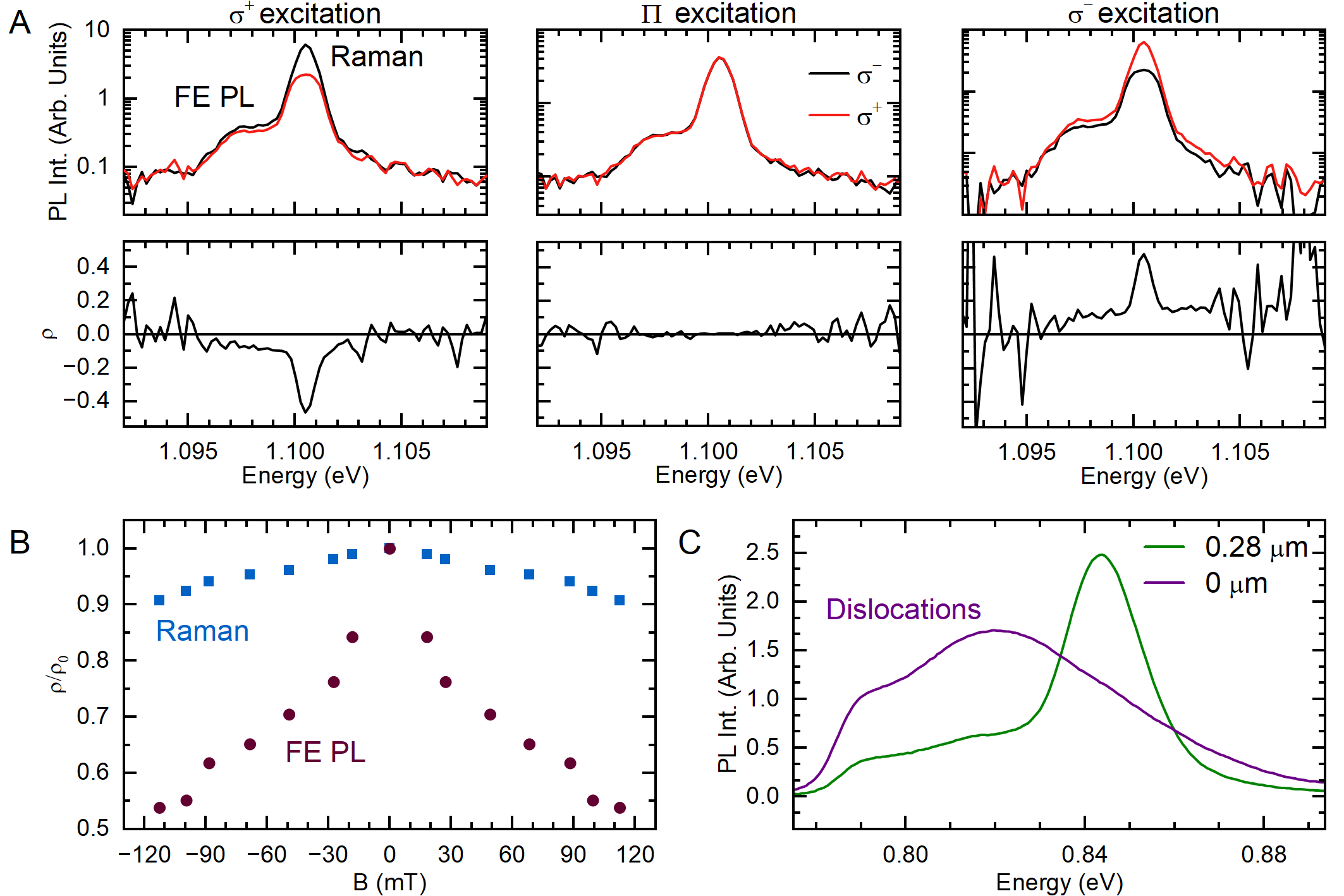}
  \caption{\textbf{Magneto-optical investigation.} \textbf{(A)} PL spectra (top) and polarization degree $\rho$ (bottom) measured by changing the helicity of the incident light. For each PL spectrum, the black (red) line corresponds to the $\sigma^-$ ($\sigma^+$) component of the Si FE PL. \textbf{(B)} $\rho$ as a function of a transverse magnetic field (Hanle effect) for the Raman line (blue squares) and for the Si FE PL (purple dots). The polarization is normalized to the zero-field value ($\rho_0$) and mirrored to negative fields to highlight the consistency with the expected Hanle line shape. \textbf{(C)} PL data in the spectral region of the direct gap of Ge for the samples with total Ge thickness equal to 0.28 $\mu$m (green) and 0 $\mu$m (violet). The spectral weight at low energy, peaked at 0.82 eV, is ascribed to dislocations in Si.\cite{Drozdow76, Reiche16}}
  \label{fig4}
\end{figure*}

A closer inspection to Fig. \ref{fig3}C surprisingly shows that the complete removal of the Ge layer still yields a residual circular polarization, which, albeit small ($\rho$=-2\%), is systematically non-zero. This opens a potential loophole in the optical spin pumping process occurring in the Ge/Si heterojunction. In the following, we will thus refine and expand our PL analysis to clarify this puzzling finding and test the role of optical spin pumping in determining the circular polarization of the Si FE.

Figure \ref{fig4}A reports polarization-resolved spectra of the etched samples measured under $\sigma^+$, linear ($\Pi$), and $\sigma^-$ excitation. Even though the largest polarization occurs at the energy of the Raman mode ($\sim 50\%$), this effective polarization is clearly affected by a sizeable polarized baseline that pertains to the broader FE PL. It can be noticed that the $\sigma^-$ PL component dominates upon $\sigma^+$ excitation, while the opposite holds true when the helicity of the pump is reversed. This perfect mirror-like behavior stems from the Kramers conjugation, making a compelling case for a spin-dependent radiative recombination. In addition, Figure \ref{fig4}A consistently shows that when a $\Pi$ pump is applied, the two $\sigma$-resolved spectra become indistinguishable. Specifically, the PL does reflect a suppression of the spin polarization, in excellent agreement with the fact that no net out-of-equilibrium spin population can be expected under linear excitation. Likewise, further evidences of spin-related emission can be seen in the energy-resolved degree of the circular polarization (see Fig. \ref{fig4}A), that is, $\rho$ switches sign for $\sigma^{\pm}$ and vanishes upon $\Pi$ excitation.

To provide conclusive proof of optical spin pumping in Si, we also conducted a dedicated magneto-optical investigation and leveraged the so-called Hanle effect.\cite{Parsons69, Dyakonov_OO} This phenomenon manifests itself as a luminescence depolarization induced by a transverse magnetic field, i.e. orthogonal to the $z$ axis, and originates from the decrease of the $z$-projection of spin polarization because of the field-induced precession of the optically injected spins. \cite{Vitiello20, marie25}

Figure \ref{fig4}B shows the field dependence of $\rho$ for both the Raman and PL peaks pertaining to the Si substrate. The Raman mode undergoes only a moderate depolarization by increasing the magnetic field. This apparent variation is certainly affected by the behavior of the underlying spectral component of the PL (see also Fig. \ref{fig4}A). Indeed, the FE emission exhibits a drastic quenching, even in a weak magnetic field of just few mT. This marked difference and the observation that luminescence nicely exhibit a distinctive Hanle lineshape provide final compelling evidences of the spin-dependent nature of the polarized emission from Si. 

Interestingly, the magneto-optical investigation discloses additional information, which is instrumental in understanding the spin dynamics characterizing the Ge/Si heterostructure. The full width at half maximum ($\Delta$B) of the FE Hanle curve is given by:\cite{Parsons69, Vitiello20} 
\begin{equation}
    \Delta B=2\frac{\hbar}{g\mu_B T_S} \label{Eq.deltaB}
\end{equation}
where $\hbar$ the reduced Plank constant, $\mu_B$ the Bohr magneton, and $g$ the electron Landé factor. $T_S$ is the spin lifetime, which depends on the spin relaxation time $\tau_s$ and the carrier lifetime $\tau$ as:
\begin{equation}
    \frac{1}{T_S}=\frac{1}{\tau_s} + \frac{1}{\tau}\label{Eq. vita spin}
\end{equation}
Owning to the non-negligible circular polarization degree of the FE, it can be reasonably assumed that $\tau_s\gg\tau$, i.e., carriers recombine before completely losing their spin orientation. As a consequence, $T_S$ $\sim$ $\tau$. Equation \ref{Eq.deltaB} can be used, knowing the well-established value of the electron g-factor of Si, to estimate $\tau\sim$ 200 ps. Such a characteristic time is extraordinarily short for an indirect bandgap transition: it is several orders of magnitude shorter than the typical minority lifetime, which in lightly-doped Si wafers is in the $\sim$ $\mu$s regime. \cite{Watters56, Schroder97, Schmidt97}

A ultra-fast lifetime is unexpected and heralds the presence of extrinsic mechanisms that in the Ge/Si heterostructures significantly accelerate the carrier dynamics compared to pristine bulk Si and might ultimately contribute to the spin pumping process. Following the calculated carrier concentration profiles that show in Si a spatially-resolved recombination rate peaked within 100 nm from interface, it is reasonable to expect that the drastic shortening of the lifetime is caused by the emergence of parasitic non-radiative channels in the proximity of the heterointerface. Crystal flaws, in the form of dislocations, are known to develop in the heteroepitaxial film because of the lattice mismatch with the substrate and are thus a likely culprit for the unexpected carrier kinetics. Such extended defects possess a large recombination velocity that can efficiently drain the photogenerated carriers, thereby reducing the interband transition rate and shortening the excess carrier lifetime.\cite{Vitiello20} Although the defective Ge/Si interface has been widely studied due to its detrimental impact on the radiative recombination in Ge,\cite{Geiger14, Pezzoli14, Pezzoli16, Grzybowski11} no attention has been paid to its possible repercussions also on Si, and dedicated investigations are still awaited. 

Here, a quick glance at Figure \ref{fig4}C demonstrates that, in the low-energy range, the PL of the 0.28 $\mu m$-thick Ge film (green, solid line) is dominated by the peak at about 0.84 eV due to recombination through the Ge direct gap. \cite{Vitiello15} Interestingly, this well-resolved PL appears to be superimposed onto a weaker but broader structure. The latter openly manifests itself in the spectra of Figure \ref{fig4}C, when the Ge layer is completely removed by the etching solution and the associated direct-gap PL suppressed. This previously hidden low-energy emission is indeed well below the Si band edge, thus pointing towards defect-induced energy levels localized within the Si bandgap. Moreover, this puzzling emission peaked at about 0.82 eV is in excellent agreement with the well-established dislocation-related luminescence, the so-called D1 line. \cite{Drozdow76, Reiche16} The distinct presence of a D line in the low-energy PL spectrum of Si demonstrates that extended defects are seamlessly injected into the substrate during the deposition of the epitaxial Ge layer. Reassurance can be found in previous structural investigations, where dislocations were poised to pierce through the Ge/Si interface and extend into the underlying Si substrate. \cite{Lee1997} 

\section{Discussion and Conclusions}

The direct experimental observation of circularly polarized emission in Si has been considered so far hindered in the conventional framework of optical spin orientation by the negligible spin-injection rate and the exceedingly long carrier lifetime pertaining to indirect gap transitions.
The measurable circular polarization accounts for the dynamical evolution of the spin ensemble given by carrier recombination and intrinsic spin relaxation. These processes effectively reduce the maximum spin polarization attainable at the moment of photoexcitation, $\rho_{max}$, so that the degree of the PL at the steady state is $\rho=\frac{\rho_{max}}{(1+\tau/\tau_S)}$. A sizeable measurement of the polarization can thus be obtained whenever extrinsic means enable a strong suppression of $\tau$ compared $\tau_S$, irrespective of the initial $\rho_{max}$ value. Viable strategies to achieve this purpose include high optical excitation to enhance non-radiative Auger recombination, as very recently shown by Ref. \cite{marie25}. The magneto-optical investigation introduced in this work demonstrated the most effective approach to ensure that spins are not entirely washed out during the electron lifetime, namely dislocations, which have been shown to induce strikingly fast dynamics in the sub-nanosecond regime. This clarifies the unprecedented residual degree of circular polarization surprisingly obtained at resonant excitation under continuous-wave excitation and upon the complete removal of the Ge film (see Fig. \ref{fig3}C).

Notably, this work additionally demonstrates how to further increase the polarization of the Si PL by manipulating $\rho_{max}$, hence overcoming the strict limitations hampering the direct optical excitation of unstrained Si. Under optical spin pumping at the steady state, a giant spin accumulation in Si can arise from the large diffusive current given by the electron spins originally photogenerated in Ge. As suggested by the diffusion model developed here, electrons optically coupled to the top of the Ge valence band can transfer spin angular momentum into Si up to an impressive maximum value of $|P_S| \simeq 30$\% (see Fig. \ref{figTeo}A and B). Such spin oriented carriers can then decay radiatively, thereby emitting circularly polarized PL. 

Symmetry arguments applied to the optical selection rules of phonon-mediated transitions in Si predict an efficiency for the emission of circular polarization in unstrained Si of $\eta_{LO} =27.7$\% and $\eta_{TO} = 18.8$\% for recombination involving LO and TO phonons, respectively. The LO/TO replicas should be cross-polarized.\cite{li10,li13} 
Consequently, in complete absence of spin relaxation mechanisms, the maximum circular polarization degree of the $\mathrm{FE_{TO}}$ is expected to be $|\rho_{max}|=\eta_{TO} \times P_S = 0.188\times 0.3 \approx 6$\% when the TO/LO splitting can be resolved. Interference between the two optical phonons otherwise decreases the efficiency, which weighted for the relative luminescence intensity of the TO, $I_{TO}$, and LO, $I_{LO}$, phonons becomes $\eta_{eff}=\frac{\eta_{TO} I_{TO}+\eta_{LO}I_{LO}}{I_{TO}+I_{LO}}= 13.3$\%, \cite{sircar14}, thus yielding $|\rho_{max}|=\eta_{eff} \times P_S = 0.133\times 0.3 \approx 4$\%. As a further reassurance of the effectiveness of the optical spin pumping approach introduced here, we can notice that these expected values of $\rho_{max}$ fall satisfactorily within the experimental uncertainty of the largest $\rho$ reported in Fig. \ref{fig3}C for a Ge thickness of 0.05 $\mu m$. 
Since under this condition $\rho\simeq\rho_{max}$, we can suggest that the injection in Si of electrons that are practically thermalized is a condition that is genuinely satisfied by the proposed optical spin pumping approach and guaranteed by the band alignment established between the spin sink, i.e., Si, and the light absorbing medium acting as spin injector, i.e., Ge. This requirement is further aided by the ultrafast dynamics established by the extended defects. In essence, the polarization of the spin current funneled into Si maps directly into the PL, giving rise to a non-monotonic behavior in which an increase of $\rho$ is initially observed in Fig. \ref{fig3}C compared to the case of pristine Si at low Ge coverage, and eventually the polarization shows an asymptotic tendency to unpolarized FE emision as the Ge epilayer approaches the $\mu m$ range.
Finally, we notice that optical spin pumping allows one to devise novel methods to increase the polarized emission in Si. It can be expected that the use of an injector comprising low-dimensional Ge structures, such as quantum wells, can enlarge $\rho_{max}$ to almost unitary values thanks to confinement-induced removal of the HH/LH degeneracy. This can be used in addition to other means such as applying strain, which has been presented by theoretical works as a strategy to (i) modify $\eta$ by lifting the degeneracy of the valence band directly in Si, \cite{li13} and (ii) lengthen $\tau_S$ by suppressing intervalley spin loss mechanisms.\cite{Tang12}

\section{Methods}\label{sec12}

\textbf{Sample structure} The Ge/Si heterostructure was grown on a Si wafer using a reduced-pressure chemical vapor deposition reactor. Starting with a 100 mm (001)Si substrate, a unintentionally-doped 1.3 $\mu$m thick layer of Ge was grown using a dual-step process. An initial low-temperature (400 $\degree$C) growth of a Ge seed layer was followed by a higher temperature (625 $\degree$C) overgrowth of a thick relaxed Ge buffer layer. Cyclic annealing at 800$\degree$C were performed, yielding Ge under a residual tensile strain. For the Ge film we estimated a p-type background doping of $2 \cdot 10^{15}$ cm$^{-3}$, whereas the Si substrate is n-type with a phosphorous donor concentration of $2 \cdot 10^{15}$ cm$^{-3}$. 
We selectively etched the Ge film using a 0.1 M H$_2$O$_2$ solution with a 10 nm/min etching rate.\cite{Huygens2007} After each etching step, we monitored the thickness of the Ge film with a Veeco Dektak 8 profilometer.

\textbf{Optical investigation} PL measurements were carried out in a closed-circuit variable-temperature cryostat. The samples were excited by a circularly polarized $\mathrm{Nd:YVO_4}$ laser at 1.165 eV. The power density of the optical pump was of the order of kW/$\mathrm{cm^2}$. The PL intensity was measured using a spectrometer equipped with a (In,Ga)As array detector. The polarization state of the PL was determined using a home-made polarimeter consisting of an optical retarder followed by a linear polarizer. Details about the determination of the polarization parameters associated to the PL polarization can be found in Refs. \cite{pezzoli12, Pezzoli13}.

\section{Acknowledgments}\label{sec14}
The authors would like to acknowledge A. Sammak for the epitaxy of Ge on Si, E. Vitiello, I. D. Losciale and M. Crippa for technical assistance with the measurements. Financial support was provided by the Air Force Office of Scientific Research under award number FA8655-22-1-7050 and by the European Union’s Horizon Europe Research and Innovation Programme under agreement 101070700. Support from PNRR MUR
project PE0000023-NQSTI is also acknowledged. M.V. and D.M. acknowledge support from Italian MUR grant PRIN 2022 PNRR Integrable Thz si-based quantum cascade operation, CUP I53D23006680001 funded by the European Union - NextGenerationEU.

\end{document}